\documentclass{jfm}
\usepackage{graphicx}
\usepackage{epstopdf, epsfig}
\usepackage{caption,subcaption}
\usepackage{marvosym}
\usepackage{tikz,siunitx}
\usepackage{listings} 
\PassOptionsToPackage{rgb}{xcolor}
\usepackage{pgfplots,pgfplotstable}
\usepgfplotslibrary{polar}
\usepgfplotslibrary{groupplots}
\pgfplotsset{compat=newest}
\usepackage{epsf} 
\usepackage{adjustbox}
\usepackage{float}
\usepackage[section]{placeins} 
\usepackage{times}
\usepackage{amsmath}
\usepackage{amssymb}
\usepackage{bm}
\usepackage{ mathrsfs }
\usepackage{lscape} 
\usetikzlibrary{plotmarks}
\usepackage{lipsum} 
\usepackage{xcolor}

\newlength{\subCaptionPos}
\newlength{\subCaptionPosTwo}
\newlength{\subFigWidth}
\newlength{\subFigHeight}
\newlength{\linePlotWidth}
\definecolor{c1}{cmyk}{1.0,0.4,0,0.26}      
\definecolor{c2}{cmyk}{0,0.62,0.88,0.15}    
\definecolor{c3}{cmyk}{0,0.25,0.86,0.07}    
\definecolor{c4}{cmyk}{0.11,0.67,0,0.44}    
\definecolor{c5}{cmyk}{0.31,0,0.72,0.33}    
\definecolor{c7}{cmyk}{0.68,0.20,0,0.07}    
\definecolor{c6}{cmyk}{0,0.88,0.71,0.36}    

\newcommand{\solidLine}[2][black]{\tikz[baseline]{\draw[line width = #2, #1] (0,.5ex)--++(.5,0);}}
\newcommand{\dashedLine}[2][black]{\tikz[baseline]{\draw[densely dashed,line width = #2, #1] (0,.5ex)--++(.5,0);}} 
\newcommand{\dashDottedLine}[2][black]{\tikz[baseline]{\draw[dashdotted,line width = #2, #1] (0,.5ex)--++(.5,0);}}
\newcommand{\dottedLine}[2][black]{\tikz[baseline]{\draw[densely dotted,line width = #2, #1] (0,.5ex)--++(.5,0);}}

\newcommand{\dott}[1]{\raisebox{2.25pt}{\tikz[baseline]{\node[draw,scale = 0.4 ,circle,fill= #1, #1](){};}}}

\newcommand{\capCircle}[1]{\raisebox{2.25pt}{\tikz[baseline]\node[black,mark size= #1]{\pgfuseplotmark{o}};}}

\shorttitle{Compressibility and variable inertia effects on heat transfer in turbulent impinging jets}
\shortauthor{J. J. Otero Perez and R. D. Sandberg}

\title{Compressibility and variable inertia effects on heat transfer in turbulent impinging jets}

\author{J. Javier Otero Perez\aff{1}
  \corresp{\email{jose.otero@unimelb.edu.au}}
  \and Richard D. Sandberg\aff{1}
 }

\affiliation{\aff{1}Department of Mechanical Engineering,University of Melbourne, Parkville VIC 3010, Australia}

\begin{document}
 
\maketitle

\begin{abstract}
{\color{black}{This article shows the importance of flow compressibility on the heat transfer in confined impinging jets, and how it is driven by both the Mach number and the wall heat-flux.}}
Hence, we present a collection of cases at several Mach numbers with different heat-flux values applied at the impingement wall.
The wall temperature scales linearly with the imposed heat-flux and the adiabatic wall temperature is found to be purely governed by the flow compression.
Especially for high heat-flux values, the non-constant wall temperature induces considerable differences in the thermal conductivity of the fluid. 
This phenomenon has to date not been discussed and it strongly modulates the Nusselt number. 
In contrast, the heat transfer coefficient is independent of the varying thermal properties of the fluid and the wall heat-flux. 
Furthermore, we introduce the impingement efficiency, which highlights the areas of the wall where the temperature is influenced by compressibility effects. 
This parameter shows how the contribution of the flow compression to raising the wall temperature becomes more dominant as the heat-flux decreases. 
Thus, knowing the adiabatic wall temperature is indispensable for obtaining the correct heat transfer coefficient when low heat-flux values are used, even at low Mach numbers. 
{\color{black}{Lastly, a detailed analysis of the dilatation field also shows how the compressibility effects only affect the heat transfer in the vicinity of the stagnation point. 
These compressibility effects decay rapidly further away from the flow impingement, and the density changes along the developing boundary layer are caused instead by variable inertia effects.
}}

\end{abstract}


\section{Introduction}
\label{sec:introduction}

Motivated by the low number of compressible flow studies investigating impinging jets, this article shows how some previously neglected fundamental compressible flow phenomena can play a significant role in achieving a correct prediction of the heat transfer at the impingement wall, even at low Mach numbers.
We focus on the underlying flow mechanisms of compressible origin and how the heat-flux prescribed at the impingement wall dictates the relevance of the compressible effects over the heat transfer.
Due to the wide range of applications that use impinging jet flows for either surface heating or cooling purposes, there has been a considerable amount of research on this type of flows over the past five decades, {\color{black}{in which the compressibility effects were not taken into account}}. 
Some of the most relevant early studies have been summarised in various reviews, such as \cite{gauntner1970survey}, \cite{jambunathan1992review} or \cite{viskanta1993heat}.
With turbomachinery applications as their principal motivation and limited by the scarce computing resources, these first studies were conducted experimentally with relatively high heat-flux values at the impingement wall.
Also, restricted by the available experimental techniques at the time, their main focus was on obtaining Nusselt number distributions at the impingement wall for different nozzle geometries, Reynolds numbers ($\Rey_D=\rho D U_b/\mu$)  and nozzle-to-plate ratios $H/D$, where $D$ is the nozzle diameter.
The Nusselt number is the non-dimensional ratio of convective and conductive heat transfer normal to the wall and is defined as 
\begin{equation}
  Nu=\frac{hD}{\kappa},
  \label{eq:Nu}
\end{equation}
where $h$ is the convective heat transfer coefficient and $\kappa$ as the thermal conductivity of the flow. 
The convective heat transfer coefficient is defined as $h = q_{w}/\left(T_{w}-T_{ref}\right)$, with $q_w$ denoting the heat-flux at the wall and $T_{ref}$ and $T_{w}$ as a reference temperature and the temperature at the wall, respectively.
The implications of the choice of $T_{ref}$ have previously been investigated mainly for unconfined impinging jets, where external factors such as the ambient temperature can play a significant role in the resulting $Nu$ profile.
The studies by \cite{goldstein1990effect} and \cite{baughn1991experimental} analysed such entrainment temperature effects on impinging jets, where they concluded that using the adiabatic wall temperature $T_{aw}$  as the reference is essential to remove these external effects and achieve a correct heat transfer coefficient at the wall.
Since the adiabatic wall temperature depends on multiple parameters such as the nozzle-to-plate ratio or the ambient temperature, these investigations introduced the concept of impingement effectiveness to non-dimensionalise $T_{aw}$.
As highlighted by \cite{viskanta1993heat}, another parameter that has an impact on the adiabatic wall temperature is the jet velocity, where the author suggests that for low jet velocities the adiabatic wall temperature is equal to the jet temperature.
In this article, we show how such an assumption might lead to significant errors in $Nu$, even at low jet velocities when the compressibility effects are minimal.  

With the improvement of experimental techniques, researchers have now become able to delve deeper into explaining the flow phenomena that shape the $Nu$ curve.
The recent experimental studies from \cite{el2012experimental} and \cite{VIOLATO201222}  at $\Rey_D \leq 5,000$ could accurately measure the time-resolved vortex dynamics that evolve along the jet's shear layer, but neither of these investigations focused on compressible flow phenomena.
At a higher Reynolds number regime ($\Rey_D = 60,000$), \cite{grenson2016investigation} also identified the coherent flow structures developing along the jet. Differently from the previous experimental investigations, they accounted for the changes in the thermal conductivity of the fluid $\kappa$ at the impingement plate as a function of temperature.
These variations in $\kappa$ with temperature are often neglected by both experimental and computational studies on impinging jet flows, and later on, we show how such simplification of the flow can lead to substantial differences in the resulting $Nu$.
In fact, the results reported by \cite{grenson2016investigation} presented significant differences with other similar studies, but they attributed this disagreement to the different flow conditions at the nozzle.
Leveraging the higher level of detail obtained from computational simulations and the greater computing power available, researchers used LES-generated datasets to unveil more details of the interaction amongst the vortical structures and the heat transfer at the wall. 
Unfortunately, the coarse resolutions used for these numerical experiments led to unclear conclusions, such as the origin of the secondary Nusselt number peak at $r/D\approx 2$ for impinging jets with $H/D<4$ \citep{hadvziabdic2008vortical,jefferson2011wall,uddin2013simulations,dairay2014turbulent}. 
This debate on such local Nusselt number enhancement was clarified by \cite{dairay2015direct}, where their DNS results showed that the constant separation of the boundary layer at that location was responsible for the local Nusselt number peak. 
These findings were in agreement with the LES results from \cite{aillaud2016secondary}, which was the first study to use a compressible flow solver ($Ma=0.1$ {\color{black}{with constant molecular viscosity $\mu$ and thermal conductivity $\kappa$}}) at a relatively high Reynolds number ($\Rey_D = 23,000$).
For a nozzle-to-plate distance of $H/D=5$, \cite{wilke2017statistics} reported the first parametric DNS study on turbulent impinging jets which investigated Mach number effects on this type of flows {\color{black}{(also with constant $\mu$ and $\kappa$)}}.
In their results for their cases at $\Rey_D=3,300$, these compressible flow effects were shown to have a strong influence in the vicinity of the flow impingement or also commonly referred to as the `jet deflection zone' \citep{gauntner1970survey} or `stagnation region' \citep{viskanta1993heat}.
In particular, they observed how the Nusselt number rises with increasing Mach number, where the cause to this phenomenon was attributed to the larger fluctuations present in the flow.
In contrast to the aforementioned compressible numerical investigations, \cite{grenson2017large} used a compressible LES code where, despite their low Mach number ($Ma \approx 0.064$), the molecular viscosity $\mu$ and the thermal conductivity $\kappa$ of the fluid were not assumed constant and varied as a function of the temperature.
For such a low Mach number setup, it seems an a-priori a reasonable assumption to neglect these variations in the flow properties \citep[such as in][]{aillaud2016secondary,wilke2017statistics}, but in this article we show how there are other factors to be considered in addition to the Mach number, which indicate the relevance of flow compressibility.
For example, for impinging jet setups with high values of $q_w$, the temperature differences between the jet and the wall will be such that the local values in the thermal conductivity of the flow will be significantly different.
Hence, modelling these setups as a full compressible flow and accounting for the variations in the flow properties as a function of temperature is necessary to obtain the correct heat transfer prediction at the impingement wall.
Note that the majority of the published literature on impinging jets assume $\kappa$ as constant.
Retrieving the fundamental definition of the Nusselt number \citep[e.g.][]{cengel2014heat}, this parameter shows the enhancement of heat transfer due to convection relative to conduction on the same fluid, where a Nusselt number of $Nu=1$ indicates pure conduction.
Thus, computing the Nusselt number from equation (\ref{eq:Nu}) using a constant $\kappa$ no longer represents the ratio of the local convection relative to conduction, but instead a non-dimensionalised heat transfer coefficient.
In this article, such parameter will be referred to as $Nu^*$. 
The data investigated herein were generated with a compressible LES solver, where the heat-flux at the wall and the thermal conductivity are defined as
\refstepcounter{equation}
$$
  q_w = \kappa_w \left. \frac{\partial T}{\partial y} \right|_w, \quad
  \kappa = \frac{\mu \left(T\right)}{RePrMa^2\left(\gamma -1 \right)},
  \eqno{(\theequation{\mathit{a},\mathit{b}})}\label{eq:q_and_k}
$$
where $T$ is the fluid temperature, $y$ is the wall normal direction and $\gamma$ is the heat capacity ratio and is assumed constant to 1.4.
{\color{black} {To further generalise our results, the heat-flux has been non-dimensionalised with the fluid properties at the jet exit, which makes it dependent on the integral flow quantities shown in (\ref{eq:q_and_k}$\mathit{b}$).
}}
Hence, assuming non-constant $\kappa$ for both convective and conductive terms for the definition of Nusselt number, equation (\ref{eq:Nu}) can be simplified to
\begin{equation}
Nu = \frac{ D }{\left(T_{w}-T_{ref}\right)} \cdot \left.\frac{\partial T}{\partial y} \right|_w.
\label{eq:Nu2}
\end{equation}

This definition is the same as used by the numerical work of \cite{dairay2015direct} (incompressible) and \cite{wilke2017statistics} (compressible), where both used a constant $\kappa$.
To date, all the published work on compressible numerical studies on impinging jets prescribed an isothermal boundary condition at the impingement wall.
As discussed earlier in this section, the use of a non-constant temperature condition (i.e. constant heat-flux) would require an additional computation using $q_{w}=0$ to obtain the adiabatic wall temperature, which is necessary for the correct evaluation of the heat transfer coefficient.
Hence, another aim of this publication is to extend the previous work of \cite{viskanta1993heat} and quantify the error when $T_{ref}$ is approximated as $T_{jet}$ for confined impinging jets, and how to recover the adiabatic wall temperature without necessarily simulating a case with $q_w=0$.
Thus, to evaluate the validity of such approximations, our study presents the first set of compressible numerical simulations of impinging jet flows with non-constant molecular viscosity, on a non-isothermal setup at various $Ma$ and $q_w$.

\section{Numerical setup}
\label{sec:numericalSetup}

The software used to generate the dataset analysed in this article is an in-house code \citep{sandberg2015compressible}, which solves the full non-linear three-dimensional compressible Navier--Stokes equations in generalised Cartesian coordinates.
In its LES mode, the software uses a fourth-order central finite difference scheme, using explicit one-sided stencils at the boundaries of corresponding accuracy \citep{carpenter1999stable}.
The solution is advanced in time with an explicit fourth-order accurate Runge--Kutta (RK) method \citep{kennedy2000low}.
To enhance the stability of the numerical scheme, a skew-symmetric splitting of the non-linear terms is applied \citep{kennedy2008reduced}. 
Also, an explicit filter is applied after every RK cycle---with a small weighting of 0.2---to remove possible spurious high-wavenumber oscillations \citep{bogey2009shock}.
To model the contribution from the unresolved scales, the code uses the WALE sub-grid scale model \citep{nicoud1999subgrid} with the standard coefficient of 0.325.
{\color{black}{The turbulent heat-flux model used in these calculations computes the diffusivity of the sub-grid scales assuming a constant turbulent Prandtl number of $\Pran_t=0.9$. 
  This numerical setup has been used successfully in other flow configurations dealing with compressibility effects \citep[e.g.][]{leggett2018loss}.
  For further detail on the validation of the current numerical setup, the reader is referred to the appendix \ref{appA}.
}}

{\setlength{\linePlotWidth}{1.5pt} 
  \begin{figure} 
    \begin{subfigure}[b]{0.525\textwidth}
      \subcaption{}
      \vspace{\subCaptionPosTwo} 
      \includegraphics{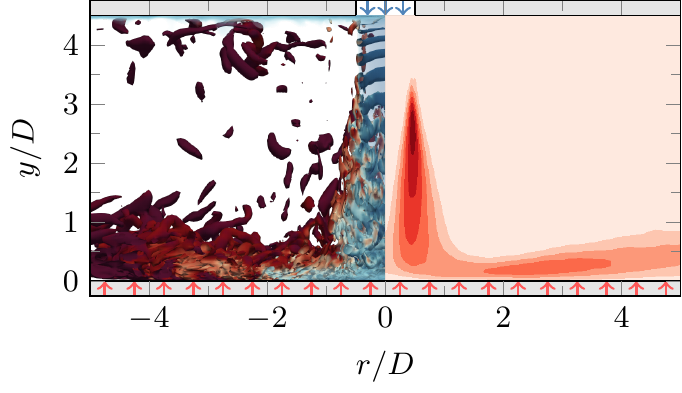} \label{fig:musgs}
    \end{subfigure} 
    \begin{subfigure}[b]{0.1\textwidth} 
      \includegraphics{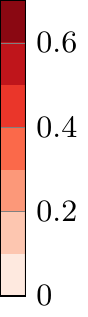} 
      \vspace{0.775cm} 
    \end{subfigure}
    \begin{subfigure}[b]{0.375\textwidth}
      \flushright
      \subcaption{}
      \vspace{\subCaptionPos}
      \includegraphics{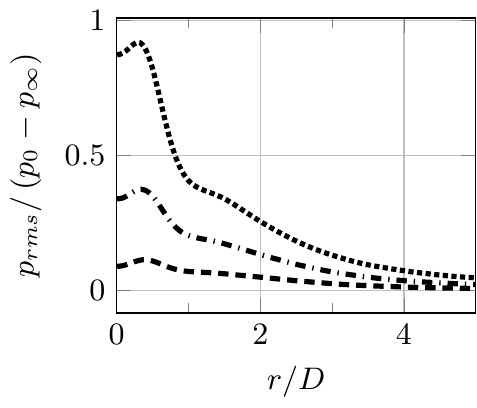} \label{fig:prms}
    \end{subfigure}   
    \caption{{\it{(a)}} Instantaneous isocontours of $Q$ coloured by temperature (left half) and time-averaged ratio of the sub-grid-scale viscosity $\mu_{sgs}$ relative to the molecular viscosity $\mu$ (right half). The red arrows along $y/D=0$ represent the constant heat-flux $q_w$ applied through that boundary. {\it{(b)}} Normalised r.m.s. of pressure at the impingement wall with $q_w=0.025$ for the three different Mach numbers: \protect\dashedLine{\linePlotWidth}, $Ma=0.3$; \protect\dashDottedLine{\linePlotWidth}, $Ma=0.5$; \protect\dottedLine{\linePlotWidth}, $Ma=0.7$.} \label{fig:musgs_and_prms}
  \end{figure}
} 

As shown in figure \ref{fig:musgs}, the coordinate system is arranged such that the $y$ direction is wall-normal, whereas the $x$ and $z$ directions are perpendicular to the jet.
The nozzle-to-plate distance is kept constant to $H/D=4.5$ for all the cases presented.
At the top wall, the jet is defined as a steady {\color{black}{circular}} top-hat profile boundary condition which is smoothed over the outer $10\%$ of the diameter with a fifth-order polynomial to prevent spurious oscillations resulting from differentiating a discontinuity.
{\color{black}{The polynomial coefficients have been chosen to produce a smooth ramp-function ($y_{ramp} \in \left[0,1\right]$ if $x_{ramp} \in \left[0,1\right]$), where the derivatives up to second order are zero at the extrema of the ramping interval $x_{ramp}$.
  This leads to a similar ramping behaviour to the functions used in other relevant investigations dealing with jet flows \citep[e.g.][]{freund2001noise,dairay2015direct,wilke2017statistics}.
}}
The jet is confined by an isothermal top wall, with the same temperature as the jet exit temperature. 
On the other hand, the impingement wall is defined as a constant heat-flux boundary, where the heat-flux is ramped down to $q_w=0$ on the grid points beyond $r/D=5$ from the stagnation point.
In contrast to other numerical investigations with this type of boundary conditions, we cannot simply prescribe a constant temperature gradient at the wall due to the non-constant character of the fluid's thermal conductivity.
Instead, as shown in equation (\ref{eq:q_and_k}), $\kappa_w$ is evaluated at each grid point along the wall for each time-step, and the temperature gradient is adjusted accordingly to match the prescribed heat-flux at the wall.
{\color{black}{From equation (\ref{eq:q_and_k}$\mathit{b}$), we see how this is due to the non-constant molecular viscosity of the fluid $\mu$, which varies as a function of temperature following Sutherland's law \citep[e.g.][]{white1991viscous}.
  All the data analysed in this article were generated using the ratio of the Sutherland's constant to the jet's temperature as 0.36867. 
  Also, the molecular Prandtl number was kept constant at $\Pran=0.7$ throughout all the cases.
  The different combinations of non-dimensional heat-flux at the wall $q_w$ and Mach number used for each case are detailed in \S\ref{sec:htScaling}.
}}

The domain spans $45D$ and $10D$ along the $x$ and $z$ directions, respectively.
At the boundaries in $x$, we use zonal non-reflecting boundary conditions of characteristic type \citep{sandberg2006nonreflecting}, whereas the domain is defined as periodic along $z$.
In this investigation, we focus on the data along the domain's centreline (i.e. $z=0$), where the data were confirmed to be independent from the distance to the periodic boundaries in a preliminary study.
The grid used for these simulations was designed iteratively to tailor the requirements of this flow at $\Rey_D=10,000$, achieving an almost constant resolution of $\Delta x^+ = \Delta z^+ \approx 40$ and $\Delta y^+ \approx 1$ within $r/D<5$ from the stagnation point.
This discretisation required $404\times110\times233$ grid points ($N_x \times N_y \times N_z$).
Despite the existing scepticism in LES data arising from the ambiguous results discussed in \S\ref{sec:introduction}, and the poor predictions achieved with some models in capturing the correct heat transfer at the wall as reported by \cite{dairay2014turbulent}, we later show in \S\ref{sec:htScaling} how our LES setup virtually matches the predictions from a DNS of the same setup. This reference DNS was discretised with over 540 million grid points and used the above numerical setup upgraded to be eighth-order accurate. 
This discretisation led to an almost constant resolution at the wall of $\Delta x^+ = \Delta z^+ \approx 8$ and $\Delta y^+ \approx 1$ within $r/D<5$ from the stagnation point.
Note that this is a finer resolution than used in \cite{dairay2015direct} for a similar $\Rey_D$.
To illustrate the contribution of the LES model in our setup, figure \ref{fig:musgs} shows the ratio of the sub-grid-scale viscosity $\mu_{sgs}$ relative to the molecular viscosity $\mu${\color{black}{ for the case with the highest model contribution ($Ma=0.3$ and $q_w=0.025$.)}}.
With the exception of the jet's shear layer at $y/D \approx 2.5$, the contribution of the model to the overall viscosity is well below $0.4$, which is an indication of a good quality LES.

\section{Temperature and heat transfer scaling at the impingement wall}
\label{sec:htScaling}

\begin{figure}
  {\setlength{\linePlotWidth}{1.5pt} 
   \setlength{\subFigHeight}{3cm}
    \begin{subfigure}[b]{0.5\textwidth} 
      \subcaption{}
      \vspace{\subCaptionPosTwo} 
      \includegraphics{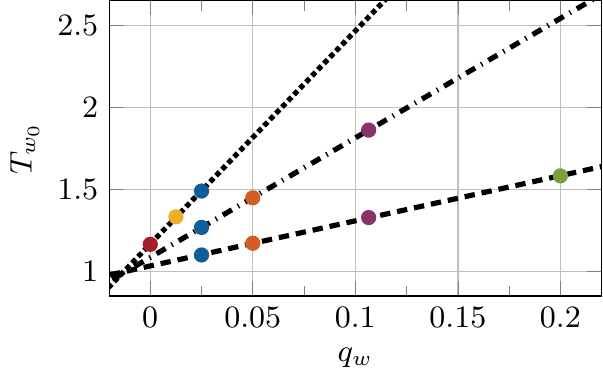}
      \vspace{0.1cm}
      \label{fig:T0_Q_M}
    \end{subfigure} 
    \begin{subfigure}[b]{0.5\textwidth}
      \flushright
      \subcaption{} 
      \vspace{\subCaptionPosTwo}
      \includegraphics{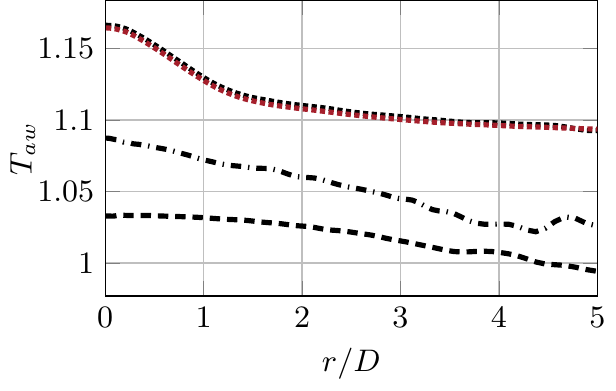}  
      \label{fig:Taw}
    \end{subfigure}
    \caption{{\it{(a)}} Temperature scaling at the stagnation point as a function of the heat-flux at the wall. Each symbol represents a different simulation. Symbols with the same colour indicate cases with the same non-dimensional $q_w$. {\it{(b)}} Linear estimation of the adiabatic wall temperature. Different line styles represent different Mach numbers: \protect\dashedLine{\linePlotWidth}, $Ma=0.3$; \protect\dashDottedLine{\linePlotWidth}, $Ma=0.5$; \protect\dottedLine{\linePlotWidth}, $Ma=0.7$. The red dotted line (\protect\dottedLine[c6]{\linePlotWidth}) shows the temperature at the wall on the adiabatic case at $Ma=0.7$.} \label{fig:q_and_Taw}
  }
\end{figure}

The following analysis is based on the data generated by ten simulations at several values of $Ma$ and $q_w$; all using the LES setup described above.
As shown in figure \ref{fig:T0_Q_M}, the results obtained with each value of $q_w$ are represented with a different colour (\protect\dott{c6}, $q_w=0.0$; \protect\dott{c3}, $q_w=0.0125$; \protect\dott{c1}, $q_w=0.025$; \protect\dott{c2}, $q_w=0.05$; \protect\dott{c4}, $q_w=0.1065$; \protect\dott{c5}, $q_w=0.2$), whereas the results from simulations at different Mach numbers are shown with a different line style (\protect\dashedLine{\linePlotWidth}, $Ma=0.3$; \protect\dashDottedLine{\linePlotWidth}, $Ma=0.5$; \protect\dottedLine{\linePlotWidth}, $Ma=0.7$).
For simplicity, this notation is kept consistent throughout this article{\color{black}{, and the different $Ma$ and $q_w$ combinations will be referred to as M0.3Q0.025 for the case at $Ma=0.3$ and $q_w=0.025$, and so on}}.
To evaluate the sole effect of $Ma$, we ran all three different Mach numbers with the same non-dimensional heat-flux $q_w=0.025$ at the impingement wall. 
As shown in figure \ref{fig:prms}, the normalised resolved pressure fluctuations along this wall increase with the Mach number. 
{\color{black}{This trend agrees with the data presented by \cite{wilke2017statistics}---based on their set of DNSs at a lower Reynolds number---where the reported pressure fluctuations also increased with the Mach number.
As a consequence of our laminar inflow profile, these pressure fluctuations do not peak at the mean stagnation point.}}
Instead, the $p_{rms}$ distributions reach maximum values at approximately $r/D \approx 0.4$ away from the stagnation location.
On the other hand, the $p_{rms}$ levels at the stagnation point are of comparable magnitude to the maximum values.
This suggests that the nozzle configuration has a relatively small effect on the flow characteristics at the impingement for the current nozzle-to-plate distance, which agrees with the findings reported by \cite{lee2000effect}.
{\color{black}{If the nozzle configuration were to have a strong influence on the flow impingement, given that our jet exit is laminar, the pressure fluctuations would in that case show a much lower magnitude at the stagnation point compared to the maximum values.}}

\subsection{The linear relation of $T_w$ and $q_w$}

Focusing again on figure \ref{fig:T0_Q_M}, we observe how the time-averaged temperature at the stagnation point increases linearly with $q_w$ for the cases at {\color{black}{each $Ma$}}.
Such linear scaling not only applies to this particular location, but to the entirety of the wall.
This linear relation has been observed in other experimental studies such as \cite{grenson2016investigation} or \cite{vinze2016influence}, and it allows for a simple estimation of $T_{aw}$ through linear extrapolation, and also without running an additional simulation with an adiabatic wall.
{\color{black}{Figure \ref{fig:Taw} confirms the validity of the linear relation of $T_w$ and $q_w$, where, for $Ma=0.7$, the estimated full adiabatic temperature profile at the impingement wall (using only the cases at $q_w=0.025$ and $q_w=0.0125$) matches exceptionally well the wall temperature obtained directly from the M0.7Q0.0 case.
}}
Differently from unconfined jets where the ambient temperature plays a significant role in the resulting adiabatic wall temperature, in confined impinging jets, the resulting $T_{aw}$ distribution is exclusively shaped by flow compression.
As the flow impinges normally onto the wall, the local flow compression gives rise to an increment in the wall temperature that strongly depends on the Mach number.
Therefore, we can break down the total temperature increment at the wall $\Delta T_w = T_w - T_{jet}$ as $\Delta T_w = \Delta T_q + \Delta T_{comp}$; where $\Delta T_{comp} = T_{aw}-T_{jet}$, and $\Delta T_q$ is the temperature increment as a result of applying a heat-flux at the wall.
The data plotted in figure \ref{fig:Taw} shows how the adiabatic wall temperature rises with increasing Mach number, as compressibility effects gain more relevance.
Note also that the compressible phenomena decay as we move away from the stagnation point.
From the data in figure \ref{fig:q_and_Taw}, we obtain that $\Delta T_{comp} \approx 1/3 \Delta T_{w}$ at the stagnation point for the cases at $q_w=0.025$. 
Intuitively, the overall contribution of $\Delta T_{comp}$ to $\Delta T_w$ decays asymptotically as the heat-flux increases ($\Delta T_q$ increases and $\Delta T_{comp}$ is independent of $q_w$), and it appears to be independent of $Ma$.
The parametric dependence of this ratio is discussed in further detail in \S\ref{sec:compressibleEfficiency}, and we show how the validity of the approximation of $T_{ref}$ as $T_{jet}$ does not depend on the jet speed \citep[e.g.][]{viskanta1993heat}, but on the prescribed non-dimensional heat-flux at the impinging wall.

\subsection{Nusselt number scaling}

\begin{figure}
  {\setlength{\subFigWidth}{3cm}  
   \setlength{\subFigHeight}{3cm}
    \begin{subfigure}[b]{0.33\textwidth}
      \subcaption{} 
      \vspace{\subCaptionPosTwo} 
      \includegraphics{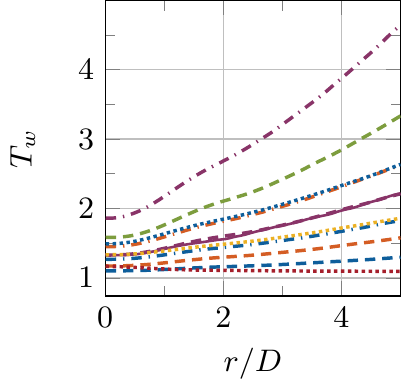}  
      \label{fig:Tw}
    \end{subfigure}
    \begin{subfigure}[b]{0.33\textwidth}
      \centering
      \subcaption{} 
      \vspace{\subCaptionPosTwo}\vspace{-0.15cm} 
      \includegraphics{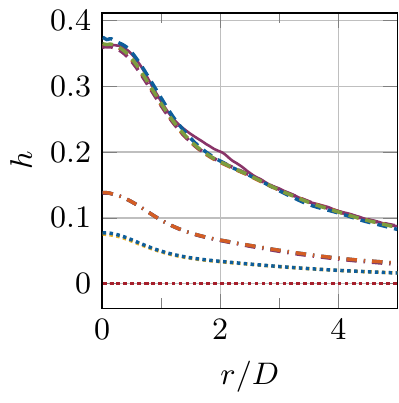}
      \label{fig:h}
    \end{subfigure}
    \begin{subfigure}[b]{0.33\textwidth}
      \flushright 
      \subcaption{}
      \vspace{\subCaptionPosTwo}
      \includegraphics{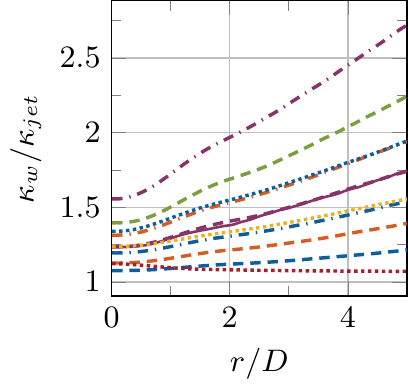}  
      \label{fig:kappa}
    \end{subfigure}
  }\\ 
  {\setlength{\subFigWidth}{9cm}
  \setlength{\subFigHeight}{4cm}
    \begin{subfigure}[b]{\textwidth}
      \centering 
      \subcaption{} 
      \vspace{\subCaptionPosTwo}
      \includegraphics{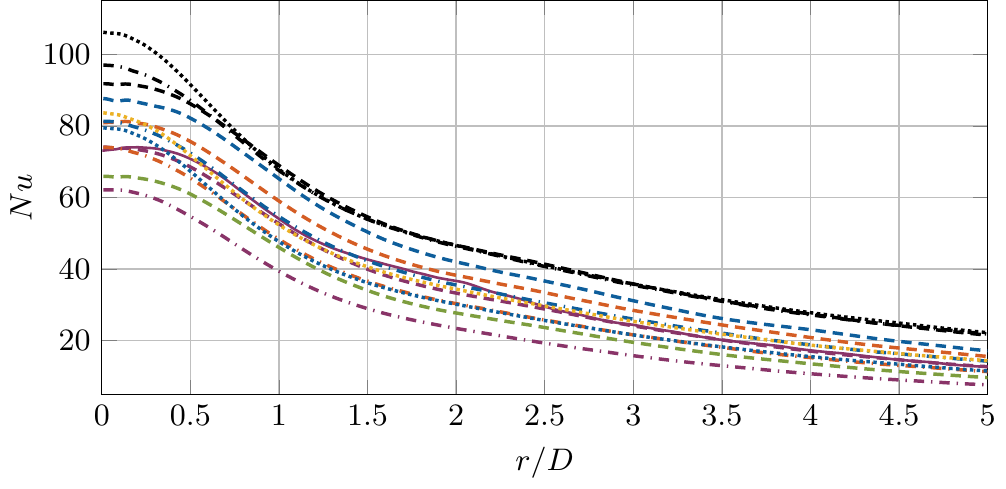}
      \label{fig:Nu}
    \end{subfigure}  
  } 
  \caption{Time-averaged temperature {\it{(a)}}, heat transfer coefficient {\it{(b)}}, thermal conductivity {\it{(c)}} and Nusselt number {\it{(d)}} distributions at the impingement wall. The solid line (\protect\solidLine[c4]{\linePlotWidth}) represents the DNS data. Other line styles indicate different Mach numbers: \protect\dashedLine{\linePlotWidth}, $Ma=0.3$; \protect\dashDottedLine{\linePlotWidth}, $Ma=0.5$; \protect\dottedLine{\linePlotWidth}, $Ma=0.7$, where different line colours indicate different $q_w$ values: \protect\dott{c6}, $q_w=0.0$; \protect\dott{c3}, $q_w=0.0125$; \protect\dott{c1}, $q_w=0.025$; \protect\dott{c2}, $q_w=0.05$; \protect\dott{c4}, $q_w=0.1065$; \protect\dott{c5}, $q_w=0.2$. {\it{(d)}} The black lines represent $Nu^*$.}
  \label{fig:Nu_breakdown}
\end{figure}

Figure \ref{fig:Nu_breakdown} breaks down the Nusselt number distribution for each case by showing the evolution of the individual terms along the impingement wall.
Also, this figure compares the performance of our LES setup at $Ma=0.3$ and $q_w=0.1065$ with DNS data at the same conditions, where our LES results show a remarkably good agreement with the DNS data.
As discussed earlier in \S\ref{sec:introduction}, LES models tend to struggle to accurately capture the phenomena that lead to the secondary Nusselt number peak for cases with $H/D<4$. 
This heat transfer enhancement takes place at $r/D \approx 2$ away from the stagnation point, which is where our DNS and LES results show the largest disagreement.
Outside this region, our LES data virtually matches the DNS results, where the errors are substantially smaller than in other previously published studies.
For example, the LES using the WALE model reported by \cite{dairay2014turbulent} overestimated $Nu$ at the stagnation point by over $25\%$ and used more than twice the number of grid points than the present setup.

Figure \ref{fig:Tw} shows the time-averaged temperature profiles at the wall for each case. 
In accordance to figure \ref{fig:T0_Q_M}, increasing the heat-flux at the wall translates to the temperature increasing linearly. 
Since the heat-flux at the wall is non-dimensionalised by the jet exit conditions, increasing the Mach number and keeping a constant $q_w$ also leads to higher wall temperature values. 
The above mentioned linear scaling between the wall temperature and the heat-flux is also reflected in figure \ref{fig:h}, where the heat transfer coefficient is independent of $q_w$ and it only varies with Mach number.
The decrease in $h$ as the Mach number is increased is also related to figure \ref{fig:T0_Q_M}, where equal changes in $q_w$ lead to larger variations in the wall temperature as the Mach number rises.
Note that the heat transfer coefficient is, in essence, the inverse of the slope governing the linear relation between $q_w$ and $T_w$ (i.e. the inverse of the slopes plotted in figure \ref{fig:T0_Q_M} give $h$ at $r/D=0$.)
At this point, it is worth mentioning that the heat transfer coefficient is calculated using $T_{ref}$ as $T_{aw}$.
As indicated by \cite{viskanta1993heat} when referring to unconfined jets, the heat transfer coefficient was also found independent from the temperature difference between the surroundings and the jet when the adiabatic wall temperature was used as the reference.
Figure \ref{fig:kappa} shows the different profiles of the thermal conductivity of the fluid at the wall normalised by the thermal conductivity at the inflow ($\kappa_{jet}$).
The distribution of $\kappa_w$ depends directly on the {\color{black}{molecular viscosity---and the temperature---of the fluid}} at the wall (see equation \ref{eq:q_and_k}), where $\kappa_w$ becomes more distant to $\kappa_{jet}$ as both heat-flux and Mach number increase (i.e. as we move further away from the incompressible flow regime).
Such variations in $\kappa_w$ {\color{black}{across the $q_w$ parameter space}} are often neglected in the literature and, as shown in figure \ref{fig:Nu}, that has a significant impact on the resulting Nusselt number.

From the data presented in figure \ref{fig:Nu}, we observe how the Nusselt number distributions decay as the heat-flux at the wall increases.
This occurs as a result of the higher conductivity of the flow (see figure \ref{fig:kappa}), which makes the convective effects less relevant to the overall heat transfer.
Also related to a higher thermal conductivity of the flow, the Nusselt number decreases with increasing Mach number, which is also due to a higher thermal conductivity of the flow.
On the other hand, an increase in $Ma$ leads to higher fluctuations in the vicinity of the stagnation point \citep[as observed by][and shown earlier in figure \ref{fig:prms}]{wilke2017statistics}, which enhances the convective heat transfer and raises the Nusselt number.
Hence, the different Mach number scaling of both conductive and convective phenomena leads to a different behaviour of the $Nu$ curves near the impingement.
In fact, in this near impingement region, the data shows how the scaling as a function of $Ma$ is strongly non-linear, where the three curves at $q_w=0.025$ show a considerable initial decay in $Nu$ from $Ma=0.3$ to $Ma=0.5$, but then it only slightly decreases again from $Ma=0.5$ to $Ma=0.7$.
Note that these Nusselt number curves have been calculated as defined earlier in equation \ref{eq:Nu2}, which is the same definition as used by \cite{dairay2015direct} (using an incompressible flow solver) and \cite{wilke2017statistics} (using a compressible flow solver).
In contrast to our findings, the Mach number study by \cite{wilke2017statistics} arrived at a different conclusion in terms of how the Nusselt number scales with $Ma$.
The origin of this disagreement arises from their results neglecting the variations in the thermal conductivity as a function of temperature.
Thus, to compare our data with all the studies which assumed constant $\kappa_w$, the black lines in figure \ref{fig:Nu} show the Nusselt number calculated as defined in equation \ref{eq:Nu} (referred to as $Nu^*$), where $\kappa=\kappa_{jet}$.
Hence, by removing the effects of the varying conductive heat transfer, the Nusselt number scaling as a function of $Ma$ looks radically different.
Close to the impingement, a higher Mach number translates into a higher $Nu^*$, which matches the observations by \cite{wilke2017statistics}.
On the other hand, away from the impingement, all three Mach number curves virtually collapse to the same values, where $Nu^*$ is independent of $q_w$.

\section{Compressible impingement efficiency} 
\label{sec:compressibleEfficiency}

\begin{figure}
  {\setlength{\subFigHeight}{3cm}
  \begin{subfigure}[b]{0.5\textwidth} 
    \caption{}
    \centering
    \vspace{\subCaptionPosTwo} 
    \includegraphics{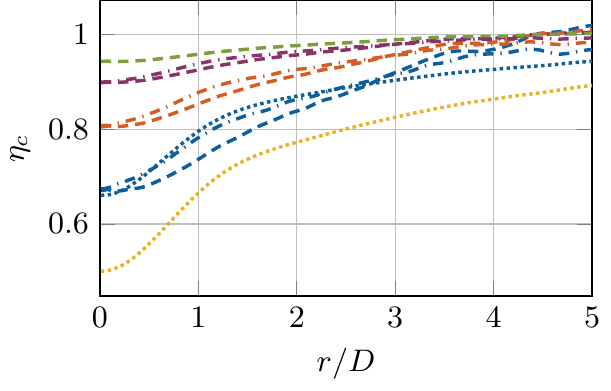}
    \label{fig:cie}
  \end{subfigure}  
  \begin{subfigure}[b]{0.5\textwidth}
    \caption{}
    \centering
    \vspace{\subCaptionPosTwo} 
    \includegraphics{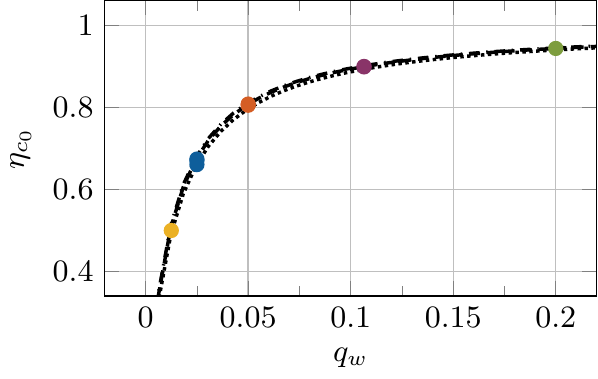}
    \vspace{0.1cm}
    \label{fig:efficiency_stagnation}
  \end{subfigure}
  \caption{{\it{(a)}} Impingement efficiency. {\it{(b)}} Impingement efficiency at the stagnation point. Different line colours indicate different $q_w$ values: \protect\dott{c3}, $q_w=0.0125$; \protect\dott{c1}, $q_w=0.025$; \protect\dott{c2}, $q_w=0.05$; \protect\dott{c4}, $q_w=0.1065$; \protect\dott{c5}, $q_w=0.2$. Different line styles show different Mach numbers: \protect\dashedLine{\linePlotWidth}, $Ma=0.3$; \protect\dashDottedLine{\linePlotWidth}, $Ma=0.5$; \protect\dottedLine{\linePlotWidth}, $Ma=0.7$.}
  \label{fig:CompAdiabaticEfficiency}
  } 
\end{figure}

As discussed earlier in \S\ref{sec:htScaling}, the adiabatic wall temperature for the present setup is exclusively governed by the flow compression which arises from the fluid impinging normally onto the wall.
Yet, to truly assess the relevance of the adiabatic wall temperature on each case, it is very common to express $T_{aw}$ as a non-dimensional parameter that allows for better comparison across different conditions.
For example, studies on unconfined impinging jets (where $T_{aw}$ is strongly modulated by the ambient temperature $T_\infty$) use the impingement effectiveness \citep{goldstein1990effect} to assess the effect of the ambient temperature on the resulting temperature distribution at the wall.
Though, for confined jet setups, the impingement effectiveness is not defined due to the lack of an ambient temperature.
Alternatively, we define the compressible impingement efficiency as 
\begin{equation}
\eta_c = \frac{T_w-T_{aw}}{T_w-T_{jet}}.
\label{eq:comImpEff}
\end{equation}
This parameter reveals the importance of the phenomena that contribute to the difference $T_{aw}-T_{jet}$ at the wall.
Hence, regions where compressibility influences $T_w$ will show $\eta_c$ values below one, whereas values of $\eta_c$ close to one suggest that it would be reasonable to assume $T_{ref}=T_{jet}$.
Note that such assumption would be extremely valuable for numerical studies, as it would save an additional computation required to obtain the adiabatic wall temperature.
Figure \ref{fig:cie} shows the impingement efficiency along the radial direction for each case.
In the vicinity of the impingement area, the drop in $\eta_c$ indicates that the wall temperature in this region is strongly influenced by compressible events.
As observed above in figure \ref{fig:Taw}, this zone exhibits the largest values in $T_{aw}$.
These compressible phenomena decay rapidly as we move further away from the stagnation point, where $\eta_c$ quickly approaches values close to one.
Despite that the flow compressibility entirely dictates the variations in $T_{aw}$, the impingement efficiency does not exhibit a large dependence on the Mach number.
Instead, $\eta_c$ appears to be dependent on the non-dimensional heat-flux at the wall $q_w$.
As shown in figure \ref{fig:efficiency_stagnation}, the impingement efficiency at the stagnation point approaches unity asymptotically as $q_w$ increases.
This asymptotic behaviour arises from the linear scaling of $T_w$ as a function of $q_w$ observed earlier in \S\ref{sec:htScaling}, where 
\begin{equation}
  \eta_c = \lim_{q_w\to\infty}  \frac{q_w \cdot m}{q_w\cdot m + T_{aw}-T_{jet}} = 1,
  \label{eq:limEff}
\end{equation}
with $m$ being the slope of the lines plotted in figure \ref{fig:T0_Q_M}.
The asymptotes for all three Mach numbers agree remarkably well, which suggests that in regions close to the impingement, the assumption of $T_{aw}$ as $T_{jet}$ does not depend on the Mach number, but on $q_w$.
In essence, figure \ref{fig:efficiency_stagnation} indicates that incompressible codes are unsuitable for accurately predicting the temperature at the wall in this setup, regardless of the imposed $q_w$.
The incompressible flow equations decouple the temperature and velocity fields under the assumption that the variations in temperature are small.
This would imply using low $q_w$ values, but as shown in figure \ref{fig:CompAdiabaticEfficiency}, the temperature distribution strongly depends on the compressible effects when $q_w$ is low.
Logically, using large $q_w$ values would reduce the weight of $T_{aw}$ over $T_w$, but the incompressible flow model is invalid for large temperature differences.
Thus, it is crucial to use a compressible code to accurately predict the temperature at the wall in the configuration studied in this article regardless of the $q_w$ used.
 
{\color{black}
\section{Compressibility and variable inertia effects} 
\label{sec:comp} 

{\setlength{\linePlotWidth}{1.5pt} 
 \setlength{\subFigWidth}{5.25cm} 
  \begin{figure} 
    \begin{subfigure}[b]{0.5\textwidth}
      \subcaption{}
      \vspace{\subCaptionPosTwo} 
      \includegraphics{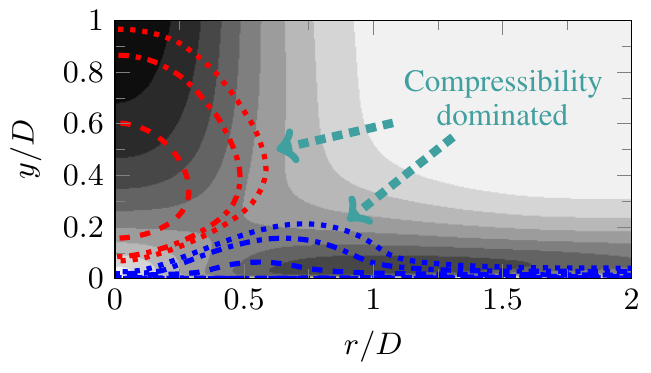} \label{fig:dilM} 
    \end{subfigure} 
    \begin{subfigure}[b]{0.5\textwidth} 
      \flushright
      \subcaption{}
      \vspace{\subCaptionPosTwo} 
      \includegraphics{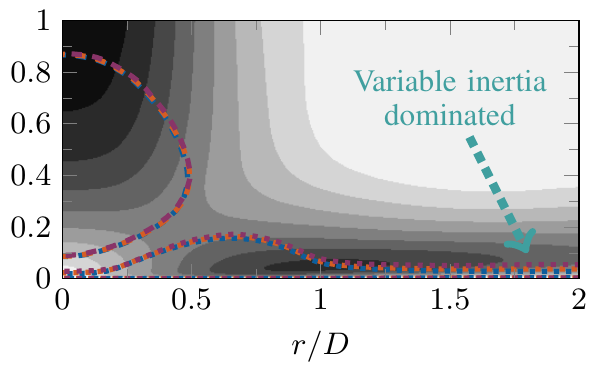} \label{fig:dilQ} 
    \end{subfigure}
    \caption{Mach number and wall heat-flux effects on the near-impingement zone. {\it{(a)}} Time-averaged contours of negative (\protect\solidLine[red]{\linePlotWidth}) and positive dilatation (\protect\solidLine[blue]{\linePlotWidth}) on cases with $q_w=0.025$. The different line styles show different Mach numbers: \protect\dashedLine{\linePlotWidth}, $Ma=0.3$; \protect\dashDottedLine{\linePlotWidth}, $Ma=0.5$; \protect\dottedLine{\linePlotWidth}, $Ma=0.7$. The background contours show the time-averaged velocity magnitude at $M=0.5$ and $q_w=0.025$. Darker contours show higher velocities. {\it{(b)}} Time-averaged contours of negative (\protect\dashedLine{\linePlotWidth}) and positive dilatation (\protect\dottedLine{\linePlotWidth}) on cases at $Ma=0.5$. Different line colours indicate different $q_w$ values: \protect\dott{c1}, $q_w=0.025$; \protect\dott{c2}, $q_w=0.05$; \protect\dott{c4}, $q_w=0.1065$. The background contours show the time-averaged velocity magnitude at $M=0.5$ and $q_w=0.1065$. The contour levels match the ones from {\it{(a)}}. The positive and negative dilatation contours in both figures show levels at $\pm0.05$.} \label{fig:dilContours}
  \end{figure}
}

In compressible flows, volume changes in the fluid elements (i.e. density variations) can occur through different physical mechanisms.
As described by \cite{lele1994compressibility}, when these volume changes are associated with changes in pressure, they are referred to as compressibility effects. 
In contrast, density variations due to heat transfer or changes in the fluid composition are known as variable inertia effects.
Given the nature of the flow setup investigated in this article, where the flow impinges normally onto a heated wall, both compressibility and variable inertia effects occur.
In the `jet deflection zone' \citep{gauntner1970survey}, the rapid decrease in the flow velocity as the flow approaches the wall leads to a flow compression.
This variation in density is unrelated to any heat transfer in the flow and changes with the Mach number, which makes it a pure compressibility effect.
Strong evidence of such a compressible phenomenon are the variations in the adiabatic wall temperature as a function of $Ma$ observed earlier in \S\ref{sec:htScaling}, and the distribution of the compressible impinging efficiency along the wall as seen in \S\ref{sec:compressibleEfficiency}.
On the other hand, variable inertia effects are expected to take place in the near-wall region due to the constant heat-flux applied through the wall, which causes an expansion of the fluid elements in that area.
However, these two physical phenomena are not mutually exclusive and they could coexist in certain regions of the flow.
Considering that the compressibility effects are associated with the Mach number and the variable inertia effects are related to the heat transfer in the fluid, one could show the regions in the flow where either of these phenomena play a role by observing the evolution of the dilatation field ($\nabla \cdot \vec{u}$) when varying $Ma$ and $q_w$ independently.

\subsection{The jet deflection zone}

{\setlength{\linePlotWidth}{1pt}
 \setlength{\subFigWidth}{2cm}
  \begin{figure}
    \begin{subfigure}[b]{\textwidth}
      \subcaption{}
      \vspace{\subCaptionPos}
      \includegraphics{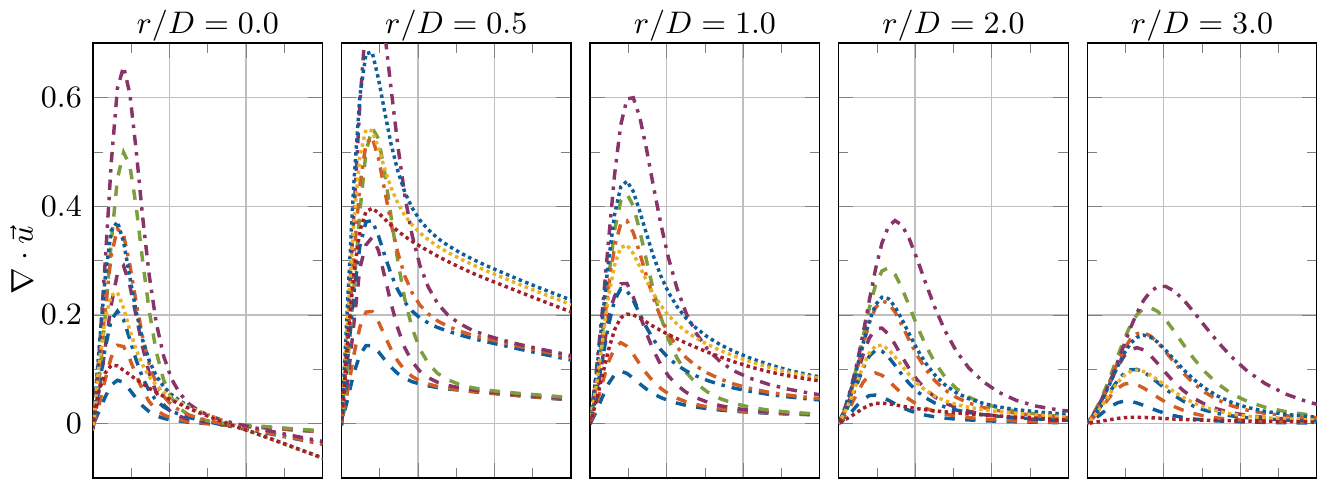} \label{fig:dilProfiles} 
    \end{subfigure}
    \begin{subfigure}[b]{\textwidth}
      \subcaption{}
      \vspace{\subCaptionPosTwo}  
      \includegraphics{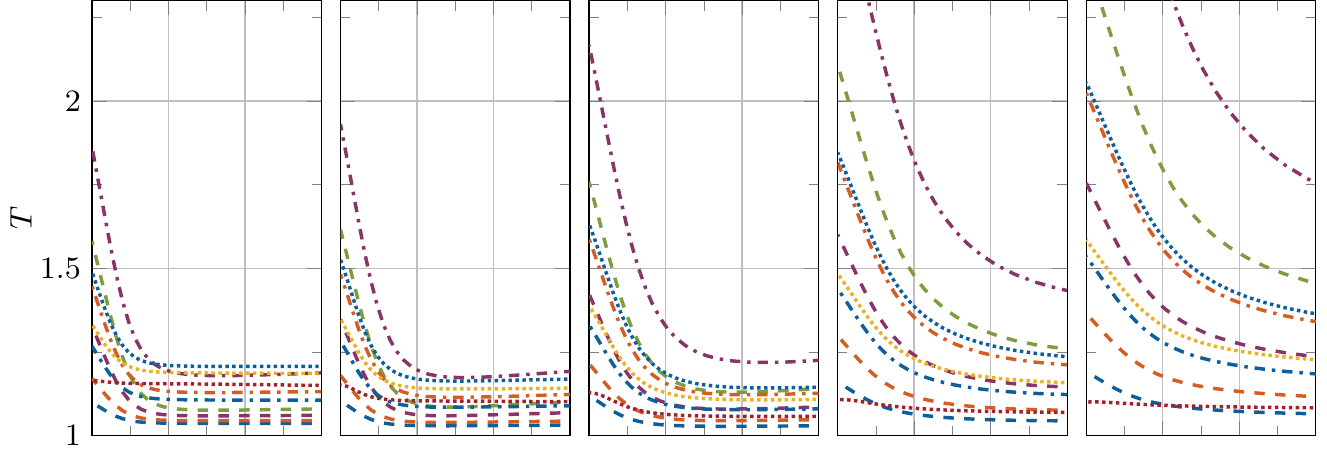} \label{fig:tempProfiles} 
    \end{subfigure}
    \begin{subfigure}[b]{\textwidth}
      \subcaption{}
      \vspace{\subCaptionPosTwo}   \hspace{-0.14cm}
      \includegraphics{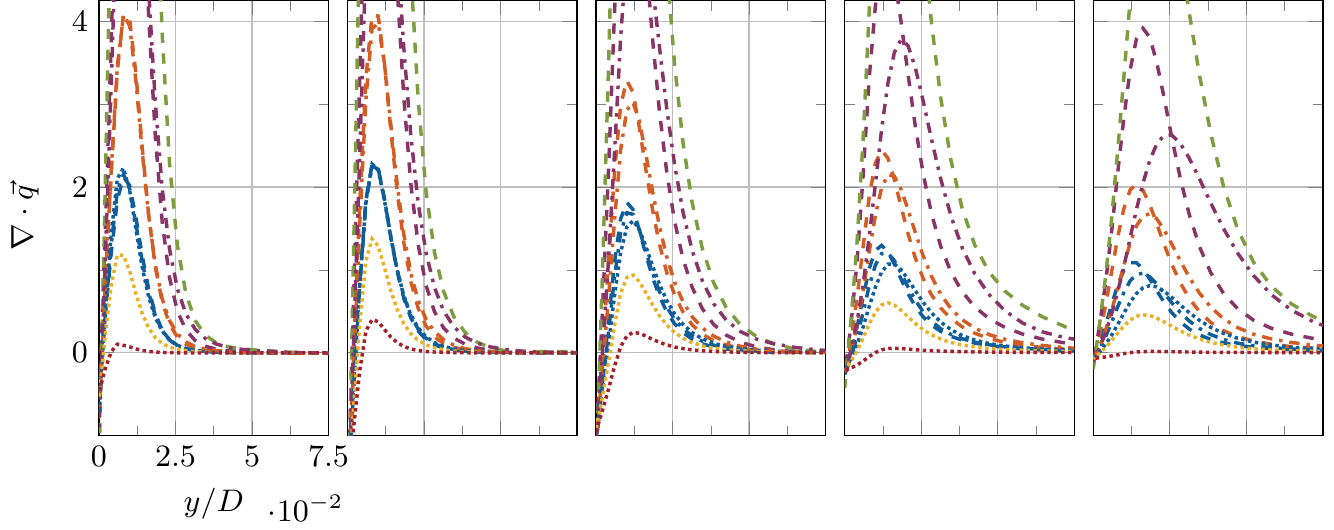} \label{fig:tempDiffusionProfiles} 
    \end{subfigure}
    \caption{{\it{(a)}} Dilatation, {\it{(b)}} temperature and {\it{(c)}} temperature diffusion profiles normal to the impingement wall at different radial locations. Different line styles show different Mach numbers: \protect\dashedLine{\linePlotWidth}, $Ma=0.3$; \protect\dashDottedLine{\linePlotWidth}, $Ma=0.5$; \protect\dottedLine{\linePlotWidth}, $Ma=0.7$. The different line colours indicate different $q_w$ values: \protect\dott{c6}, $q_w=0.0$; \protect\dott{c3}, $q_w=0.0125$; \protect\dott{c1}, $q_w=0.025$; \protect\dott{c2}, $q_w=0.05$; \protect\dott{c4}, $q_w=0.1065$; \protect\dott{c5}, $q_w=0.2$.} \label{fig:wallProfiles}
  \end{figure} 
} 

Figure \ref{fig:dilContours} shows negative and positive contours of dilatation at the near-impingement area, highlighting the evolution of the zones with flow compression and expansion as a function of the Mach number (fig. \ref{fig:dilM}) and the wall heat-flux (fig. \ref{fig:dilQ}).
In essence, the red contours in figure \ref{fig:dilM} show the locus of the jet deflection zone with flow compression and how this area grows larger with the Mach number.
Note that these variations in the dilatation field are exclusively induced by changes in $Ma$ (i.e. a pure compressibility effect), since all the three cases represented in this figure have the same $q_w$.
The widening of this flow compression zone appears to saturate as the Mach number increases, where the growth occurs mainly in the lateral and upward directions.
In the near-wall region, the temperature excess from the normal flow compression cannot be evacuated through the wall (which is also heated), and it leads to a flow expansion in the vicinity of the stagnation point. 
Such temperature build up agrees with our observations earlier in \S\ref{sec:htScaling} and \S\ref{sec:compressibleEfficiency}, where $T_{aw}$ was shown to be greater than $T_{jet}$ and dependent on $Ma$.
Since this flow expansion is entirely caused by the heat transfer at the near-wall region, it is thus a full variable inertia effect, which sets off a balancing mechanism with the counteracting normal flow compression described above.
From figure \ref{fig:dilQ}, we see how increasing the heat-flux does not cause significant changes to the flow compression region, where all the cases at $Ma=0.5$ show matching contours of negative dilatation.
This independence of the compressibility effects with respect to the heat-flux at the wall supports the observations from \S\ref{sec:htScaling}, where we showed that linearly extrapolating the adiabatic wall temperature--exclusively governed by the flow compression--from cases at other heat-fluxes is a valid approximation.
Despite the changes in the dilatation field as a function of $q_w$ are minimal, the background contours shown in figures \ref{fig:dilM} and \ref{fig:dilQ} show that the imposed heat-flux has also an effect on the velocity field.
These two figures show the time-averaged velocity field for the M0.5Q0.025 and M0.5Q0.1065 cases, where the velocity in the jet deflection zone (at $r/D\approx0.5$ and  $y/D\approx 0.2$) increases with the heat-flux at the wall.
The source of this phenomenon resides in the higher temperature of the flow surrounding the jet as $q_w$ is increased. 
As seen earlier in \S\ref{sec:numericalSetup} with the definition of $\mu\left(T\right)$, a higher flow temperature causes an increase in the local molecular viscosity.
Hence, this higher viscosity along the shear layer causes a more coherent jet impingement, which leads to the higher velocity values observed in this area.

Figure \ref{fig:wallProfiles} provides a closer look on the near-wall region, showing the wall normal profiles of dilatation, temperature and temperature diffusion at different radial locations.
Focusing on the dilatation profiles at $r/D=0.0$ (figure \ref{fig:dilProfiles}), it can be seen that, in agreement with figure \ref{fig:dilQ}, the magnitude of the dilatation in the flow compression zone shows almost no differences across different $q_w$ values. 
Instead, these curves collapse into a single line for each Mach number.
This suggests that the flow compression zone is mostly dominated by compressibility effects, where severe changes to the wall heat-flux might only introduce minor modulations in the overall flow compression.
As a matter of fact, figure \ref{fig:tempProfiles} shows how higher $q_w$ values also lead to higher temperatures even inside the compression zone, but as seen above, the sensitivity of the flow compression in that region to temperature changes is minimal.
Closer to the wall, the flow expansion in this near-wall region increases with both $q_w$ and $Ma$.
This phenomenon occurs due to a higher temperature build up in that area arising from both a higher flow compressibility and a higher heat-flux at the wall.
To give a measure of this temperature build up phenomenon, figure \ref{fig:tempDiffusionProfiles} shows the wall normal profiles of the temperature diffusion. 
There, it can be observed how each case has a negative temperature diffusion at the wall near the impingement.
This is a direct consequence of the above mentioned balancing mechanism taking place between the compressibility and variable inertia effects. 
In essence, the fluid elements located in the stagnation point area have a positive heat-flux imposed from both ends in the vertical direction.
From the top and bottom, the temperature excess from the normal flow compression and the constant heat-flux applied at the wall heat up the flow in that area and lead to a negative temperature diffusion at the wall.
Note that even the M0.7Q0.0 case shows a negative temperature diffusion in this area as the flow is unable to evacuate the temperature excess from the flow compression alone through the adiabatic boundary.
In fact, the temperature diffusion and dilatation plots are strongly related, where they share the location of their maximum values.
Thus, this shows how the nature of the boundary condition applied at that boundary introduces the variable inertia effects at the wall which are responsible for the flow expansion in this region.

\subsection{The wall jet zone}

As the fluid moves away from the stagnation region, it undergoes a rapid expansion which leads to an acceleration along the radial direction between $0 \lessapprox r/D \lessapprox 1$.
The contours of positive dilatation from figure \ref{fig:dilContours} show how this flow expansion becomes larger with the Mach number in this initial region, but these differences rapidly decay as we move further away from the impingement.
In contrast, the flow expansion contours in the near impingement ($r/D \lessapprox 1$) shown in figure \ref{fig:dilQ} do not exhibit a large sensitivity respect to the heat-flux applied at the wall. 
However, beyond this area, these contours do manifest a dependency on $q_w$ and show a visible trend; moving further off the wall as a function of the wall heat-flux. 
In essence, the evolution of the positive dilatation contours shown in figure \ref{fig:dilContours} suggests that the flow expansion in the initial wall jet region is dominated by compressibility effects (hence the dependency on $Ma$), whereas the expansion occurring in zones further away from the impingement is mostly governed by variable inertia effects (hence the dependency on $q_w$).
On the other hand, the velocity contours from figure \ref{fig:dilContours} show that the velocity field is also affected by $q_w$ in the wall jet zone, where the contours in the near-wall region at $r/D\approx 1$ are distinctly darker for the case M0.5Q0.1065 compared to those in the case M0.5Q0.025.
These differences in the near-wall velocity field are also reflected in the wall shear stress ($\tau_w = \mu_w/Re_D \cdot \left.\partial u/\partial y\right|_w$) as shown in figure \ref{fig:tWall}.
In essence, the variations in $\tau_w$ as a function of $q_w$ reveal the importance of coupling the temperature and the velocity field for these type of flows, where the differences are relevant even for the cases at the lower Mach number $Ma=0.3$.
In fact, both the heat-flux at the wall and the Mach number modulate the magnitude and the location of the peak in the wall shear stress profile ($\tau_{w_{max}}$).
Similarly to the peak of the pressure fluctuations shown earlier in figure \ref{fig:musgs_and_prms}, the location of $\tau_{w_{max}}$ suffers a shift towards the stagnation point as the Mach number is increased. 
In contrast, $\tau_{w_{max}}$ moves further away from the impingement as the heat-flux at the wall is increased.
However, the magnitude of these variations decays very rapidly along the radial direction, where eventually, all the curves collapse into a single line.

{\setlength{\subFigWidth}{9cm}
  \setlength{\subFigHeight}{4cm}
  \begin{figure} 
  \centering 
    \includegraphics{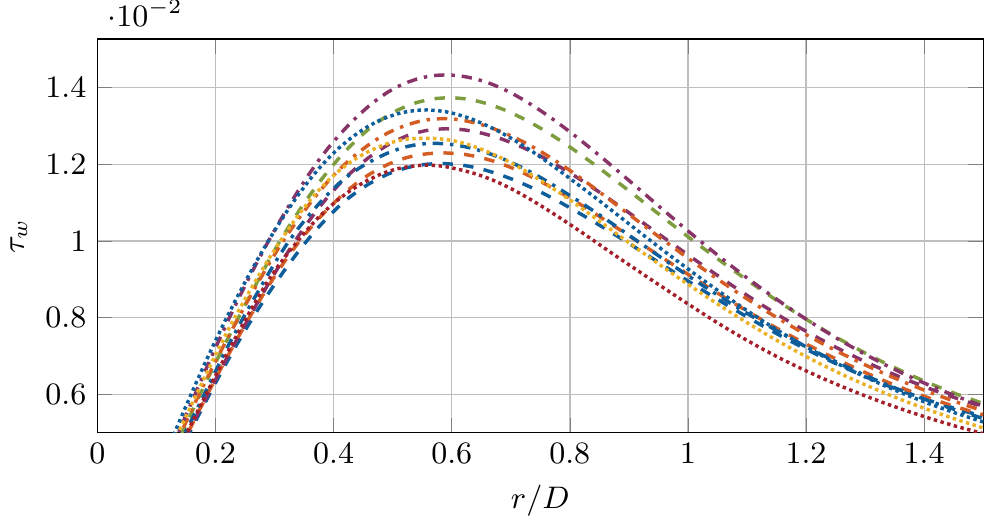}
    \caption{Time-averaged wall shear stress at the impingement wall. The different line styles indicate different Mach numbers: \protect\dashedLine{\linePlotWidth}, $Ma=0.3$; \protect\dashDottedLine{\linePlotWidth}, $Ma=0.5$; \protect\dottedLine{\linePlotWidth}, $Ma=0.7$, where different line colours indicate different $q_w$ values: \protect\dott{c6}, $q_w=0.0$; \protect\dott{c3}, $q_w=0.0125$; \protect\dott{c1}, $q_w=0.025$; \protect\dott{c2}, $q_w=0.05$; \protect\dott{c4}, $q_w=0.1065$; \protect\dott{c5}, $q_w=0.2$.}
    \label{fig:tWall}
   \end{figure} 
}

Focusing again on figure \ref{fig:dilProfiles} and exploring the density changes in the near-wall region with more detail, we observe how both the Mach number and the heat-flux at the wall have an effect on the flow expansion occurring in this area.
In the closer zones to the impingement ($r/D \lessapprox 1$), raising the Mach number or the heat-flux at the wall increases the magnitude of the flow expansion without modifying the vertical location of the peak in the dilatation profile.
Moreover, in the cases with the same Mach number, their profiles collapse into the same curve shortly after the peak and before the dilatation has decayed to zero.
Hence, this implies that at the near-wall region, both variable inertia and compressibility effects play a role, but the variable inertia effects rapidly cease to have an effect on the dilatation field as we move away from the wall.
Further away from the impingement, the compressibility effects decay and the constant $q_w$ continues to feed the variable inertia effects along the wall, which keep expanding the flow in this region.
As it can be observed in figure \ref{fig:tempProfiles}, this variable inertia phenomenon increases the thickness of the developing thermal boundary layer along the radial direction, which displaces the peak in the dilatation profile further off the wall.
On the other hand, note how, in contrast to every other case, the lack of a positive heat-flux at the wall in the case M0.7Q0.0 causes the temperature of the wall to decrease in the radial direction, as the compressibility effects decay.
This reduction of the compressibility effects is also reflected in the temperature diffusion profiles (figure \ref{fig:tempDiffusionProfiles}), where the negative values at the wall move towards zero along the radial direction. 
This trend shows that, due to the absence of the aforementioned balancing mechanism between the compressibility and variable inertia effects, the boundary layer is now able to expand and diffuse the heat away from the wall.
Lastly, as noted above when analysing the jet deflection zone, these plots in the wall jet zone also show the strong correlation between the temperature diffusion and the dilatation field, where the peak of both quantities also occurs at the same wall normal locations. 
}

\section{Conclusions}
\label{sec:conclusions}

The first set of compressible numerical simulations of impinging jet flows with non-constant molecular viscosity for a non-isothermal setup were presented.
To assess the compressibility effects over the heat transfer at the impinging wall, our dataset was generated using different values of $Ma$ and $q_w$.
This LES setup was iteratively designed to achieve a good resolution at the wall while keeping the grid size as low as possible, where a comparison with DNS data showed the LES results to be reliable.
At a constant Mach number, the temperature at the wall scaled linearly with the heat-flux $q_w$.
This permitted the estimation of the adiabatic wall temperature through a simple linear extrapolation, where $T_{aw}$ was found to be exclusively governed by the {\color{black}{compressibility effects present in the flow that arise from the normal flow impingement}}.
{\color{black}{Such approximation of $T_{aw}$ was verified for $Ma=0.7$, showing an excellent agreement with the wall temperature obtained by running a case with an adiabatic wall.}}
Similarly, the temperature at the wall also increased with $Ma$ when $q_w$ was kept constant.
All these changes in temperature induced considerable variations in the thermal conductivity of the fluid at the wall, which had a significant impact on the resulting Nusselt number.
To date, this dependency of $Nu$ on the heat-flux at the wall had never been reported in the literature, as most of the investigations neglected variations in $\kappa_w$.
On the other hand, when using the adiabatic wall temperature as the reference temperature, the heat transfer coefficient was independent of the heat-flux, and it decayed with increasing Mach number.
Through the impingement efficiency, we were able to show the areas of the wall where the {\color{black}{compressibility effects}} play a significant role in the resulting $T_w$.
Furthermore, the impingement efficiency exhibited an asymptotic behaviour approaching unity as the heat-flux at the wall was increased.
This shows that the validity of approximating $T_{ref}$ as $T_{jet}$ to calculate the Nusselt number depends on $q_w$, instead of the Mach number {\color{black}{\citep[as previously reported by, for example,][]{viskanta1993heat}}}.
Also, the impingement efficiency decreased for the lower values of $q_w$, which suggests that incompressible codes are unable to obtain an accurate prediction of the temperature at the wall for the present setup {\color{black}{regardless of the $q_w$ values used}}.
{\color{black}{Lastly, in agreement with the impingement efficiency, a detailed analysis of the dilatation field showed how the compressibility effects dominate the density variations in the vicinity of the stagnation point. 
In contrast, the density changes in areas further away from the impingement were linked instead to variable inertia effects arising from the heat-flux applied at the wall. 
At a constant $Ma$, varying the heat-flux at the wall did not cause any significant changes to the compressibility dominated areas. 
In fact, the compression region in the jet deflection zone was shown to have almost no sensitivity to changes in the temperature of the flow.
However, mainly in the wall jet zone, the velocity field did show to be affected by the changes in temperature induced by higher $q_w$ values.
These variations in the flow topology were also reflected in the wall shear stress along the impingement wall, which suggests that the coupling of the temperature and velocity fields plays an important role even at low Mach numbers.
}}

\medskip
This work was supported by a grant from the Swiss National Supercomputing Centre (CSCS) under project ID s877 and by resources provided by the Pawsey Supercomputing Centre with funding from the Australian Government and the Government of Western Australia.

\medskip
Declaration of Interests. None.
  
\begin{figure} 
\centering 
  {\setlength{\subFigWidth}{9cm} 
  \setlength{\subFigHeight}{4cm}  
  \setlength{\linePlotWidth}{1.5pt}
      \includegraphics{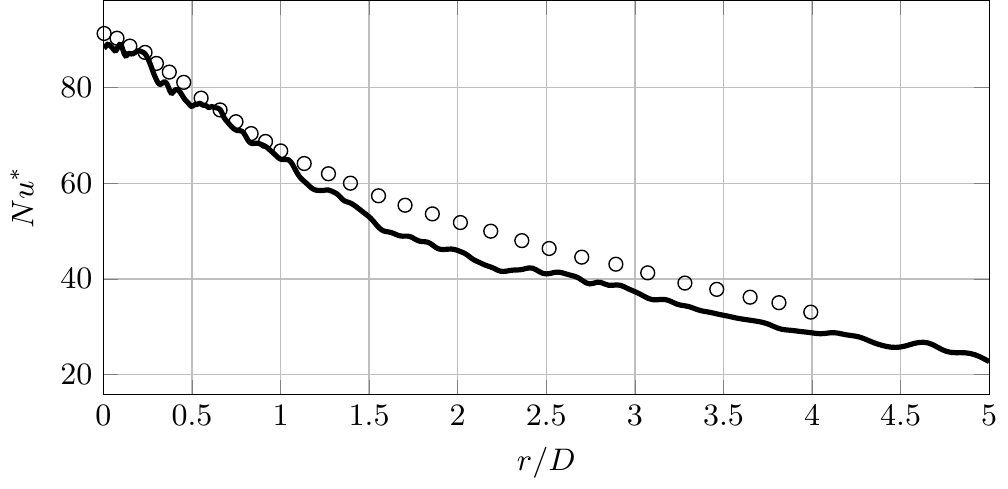}
      \label{fig:NuVal}
  }
  \caption{Time-averaged $Nu^*$ for an impinging jet setup at $H/D=6$ and $\Rey=10,000$. The solid line shows the results from our validation DNS, whereas the symbols  (\protect\capCircle{2pt}) show the data from \cite{lee1999stagnation}.
   }
  \label{fig:Validation}
\end{figure}

{\color{black}
{
\appendix
\section{}\label{appA}

Prior to generating any of the data analysed in this article, a preliminary DNS was conducted to validate the impinging jet setup with our in-house code.
For simplicity, we refer to this setup as the validation setup, whereas the above-described numerical setup is referred to as the production setup.
Similarly to the production setup, this case simulated a confined circular impinging jet flow at a Reynolds number (based on the jet's bulk velocity) of $\Rey=10,000$ and at a Mach number of $Ma=0.3$.
The boundary conditions used for the validation case match the ones used for the production setup, with the exception of the impingement wall, which is set to a no-slip isothermal wall with $T_w = 1.3T_{jet}$.
On the other hand, the domain dimensions differ from the production setup, where in this case the nozzle-to-plate distance was set to $H/D=6$ to replicate the experimental reference case \citep{lee1999stagnation}.
Along the $x$ and $z$ directions, the domain spans $24D$ and $10D$, respectively; where the jet is located at the centre of the domain.
The domain was discretised with $832 \times 750 \times 696$ grid points ($N_x \times N_y \times N_z$), leading to a grid resolution at the impingement wall of $y^+ \lesssim 0.71$ (maximum $y^+$ values occur at $0.65D$ away from the impingement point). In the directions parallel to the wall, both $x^+$ and $z^+$ exhibit values which range from 10 to 11.4 between $1D$ and $3.5D$ away from the impingement location. Outside this region, both $x^+$ and $z^+$ drop below 6.
Figure \ref{fig:Validation} compares the $Nu^*$ resulting from our validation setup with the experimental data reported in \cite{lee1999stagnation}.
This experiment was also conducted with round impinging jets at $\Rey=10,000$ and $H/D=6$.
Despite that \cite{lee1999stagnation} used a fully turbulent pipe to generate the jet flow and our simulation used a fixed profile as described in \S\ref{sec:numericalSetup}, our DNS data shows a good agreement with their results, which validates our DNS setup.
Hence, with the DNS setup validated, the good agreement between the LES and DNS results shown in \S\ref{sec:htScaling} also validates our LES setup.
}}

\bibliographystyle{jfm}   
\bibliography{Bibliography.bib}

\end{document}